\documentstyle[sprocl,epsfig]{article}

\def\Journal#1#2#3#4{{#1} {\bf #2}, #3 (#4)}

\def\PLB{{\em Phys. Lett.}  B}

\def\apj{{\em Astroph. J.}}
\def\aa{{\em Astron. Astroph.}}
\def\mnras{{\em M.N.R.A.S.}}
\def\pasp{{\em P.A.S.P}}

\def\be{\begin{equation}}
\def\ee{\end{equation}}
\def\bea{\begin{eqnarray}}
\def\eea{\end{eqnarray}}

\begin{document}

\title{JETS AT ALL SCALES}

\author{F. TAVECCHIO}

\address{INAF-Osservatorio Astronomico di Brera-Merate \\E-mail:
tavecchio@merate.mi.astro.it}  

\maketitle\abstracts{I discuss recent developments in the field of
relativistic jets in AGNs. After a brief review of our current knowledge
of emission from Blazars, I discuss some consequences of the recent
detection made by {\it Chandra} of X-ray emission from extended
jets. Finally I report some recent results on the problem of the
connection between accretion and jets, study that in principle could shed
light on the important issue of jet formation.}

\section{Introduction}

Jets are ``pipelines'' through which energy and matter originating from
the central region of AGNs can flow out from the galaxy and, as in FRII
radio galaxies, reach the radio-lobes (e.g. Begelman, Blandford \& Rees
1984). The comprehension of such structures represents a great challenge
for Astrophysics: the ultimate goals include the understanding of the
physical processes able to produce such energetics and collimated outlows
and the complex dynamics of the jet and its interaction with the
environment.

In the last decade, thanks to new instruments (in particular the $\gamma
$-ray telescope EGRET, the TeV Cherenkov telescopes and {\it Chandra}),
this field has received a great impulse. In the following I briefly
present our current view of the principal physical processes acting in
jets, from the innermost portion of the jet close to the central
``engine'' to the external regions. Finally I discuss some works in
progress regarding the connection between the inner portion of the jet 
and the outer regions.

\section{The inner jet: Blazars}

It is now generally accepted that the blazar "phenomenon" (highly
polarised and rapidly variable radio/optical continuum) is due to a
relativistic jet pointing close to the line of sight (e.g. Urry \&
Padovani 1995). Relativistic effects strongly amplify the emitted
non-thermal continuum produced close to the central BH (at $d<0.1$ pc,
Fig.1) that can dominate over the thermal contribution from the disk or
the Broad Line Region. Therefore the study of these sources offers us the
unique possibility to probe the jet in its initial portion.

\subsection{Spectral properties}

The typical Spectral Energy Distribution of blazars presents two broad
humps, the first peaking in the IR through X-ray band, the second in the
$\gamma $-ray domain, extending in some sources to the TeV band (Fig.1).

The high degree of polarization clearly indicate that the low energy
component is produced through synchrotron emission. The second peak is
likely produced through IC scattering of soft photons by the same
electron population responsible for the synchrotron emission (although
other mechanisms have been proposed, see e.g. the review in Sikora \&
Madejski 2001). In the {\it Synchrotron-Self Compton} model it is assumed
that the soft photon energy density is dominated by the synchrotron
photons themselves, while the so-called {\it External Compton} models
assume that soft photons coming from the external environment (disk, BLR)
dominate over other possible contributions.
\begin{figure}
\centerline{\hspace{4.5 cm}\epsfig{figure=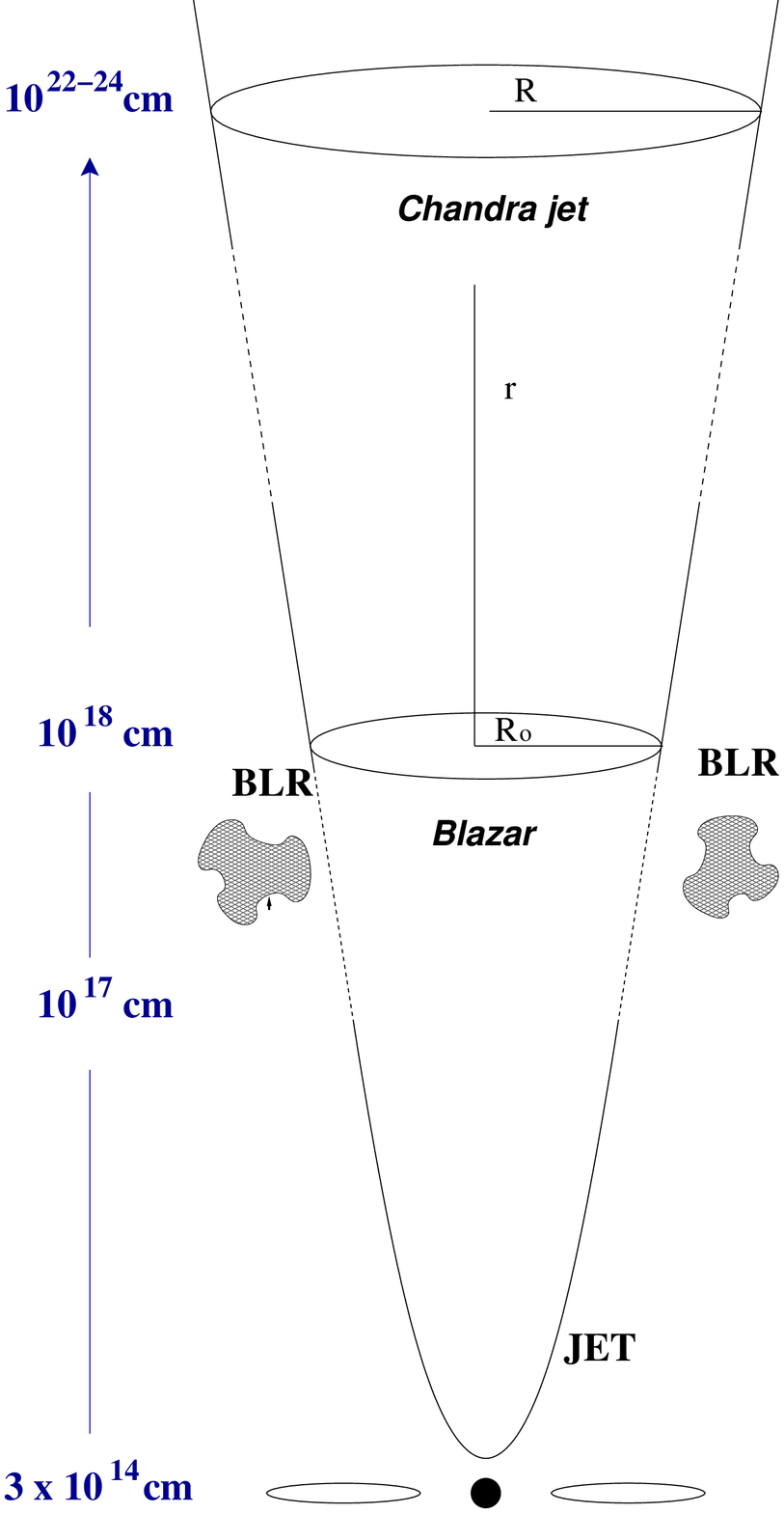, width=5.cm, height=7.5 cm}}
\caption{Sketch of the current view of extragalactic relativistic
jets. The flow is accelerated and collimated close to the BH. At a
distance $\sim 0.1$ pc part of the power is dissipated and those jets closely
aligned to the line of sight appears as Blazars. After this region the
jet propagate with small dissipation.} 
\end{figure}
\begin{figure}
\centerline{\hspace{3.5 cm}\epsfig{figure=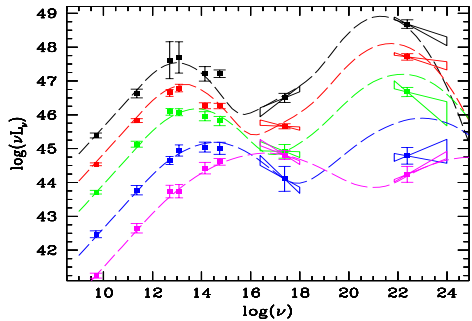, width=7.0cm,height=5.0cm}}
\caption{The blazar spectral sequence (from Fossati et al. 1998).} 
\end{figure}
Although blazars form a rather inhomogeneous class from the point of view
of the optical classification, the SEDs present a notable trend with the
luminosity, the so-called {\it blazars sequence} (Fossati et al. 1998;
Donato et al. 2001). The sources with high luminosity output ($L\sim
10^{48}$ erg cm$^{-2}$ s$^{-1}$, sources mainly classified as Flat
Spectrum Radio Quasars) have both peaks located at low energies, the
first in the IR region and the high energy peak around 1 MeV. On the
contrary, at low power the sources (BL Lac objects, showing weak or even
absent thermal features) display very high peak frequencies: in HBL,
which appear to be the less powerful objects (with typical luminosities
$10^{42}$ erg cm$^{-2}$ s$^{-1}$) the first peak falls in the UV-soft
X-ray band and the high energy component can peak near the TeV region.

This behaviour has been interpreted (Ghisellini et al. 1998, recently
revisited in Ghisellini et al. 2002) as the result of a balance between a
acceleration an cooling suffered by the electrons producing the radiation
at the peaks. In the assumption that the acceleration rate is constant in
all the sources, the sequence can be successfully interpreted as due to
an increasing cooling rate (dominated in most of the sources by IC
losses) from the low-power BL Lacs to the powerful FSRQs.

\subsection{Jet power and matter content}

SEDs can be satisfactorily reproduced by simple emission models
(e.g. Ghisellini et al. 1998, Tavecchio et al. 1998), yielding the value
of the main physical quantities of the jet, such as electron density and
energy, magnetic field, size of the region. In particular measuring the
X-ray spectra and adapting a broad band model to their SEDs yields
reliable estimates of the total number of relativistic particles
involved, which is dominated by those at the lowest energies. This is
interesting in view of a determination of the total energy flux along the
jet (e.g. Celotti et al. 1997, Sikora et al. 1997). The total " kinetic"
power of the jet can be written as:
\begin{equation}
P_{\rm jet}=\pi R^2 \beta c \,U \Gamma ^2
\end{equation}
\noindent
where $R$ is the jet radius, $\Gamma$ is the bulk Lorentz factor and $U$
is the total energy density in the jet, including radiation, magnetic
field, relativistic particles and eventually protons. If one assumes that
there is 1 (cold) proton per relativistic electron, the proton
contribution is usually dominant.

In Fig. 2 the derived radiative luminosity $L_{\rm jet}$ and kinetic
power of the jet $P_{\rm jet}$ for a group of sources with sufficiently
good spectral information are compared. The ratio between these two
quantities gives directly the ``radiative efficiency'' of the jet, which
turns out to be $\eta\simeq 0.1$, though with large scatter.  The line
traces the result of a least-squares fit: we found a slope $\sim 1.3$,
suggesting a decrease of the radiative efficiency with decreasing power.

\subsection{The accretion-jet connection}

Two main classes of models for the production of jets has been
proposed. The first class considers the extraction of rotational energy
from the black hole itself (Blandford \& Znajek 1977), while the in the
models of the other class one considers MHD winds associated with the
inner regions of accretion disks (Blandford \& Payne 1982). In both cases
the extractable power critically depends on the value of the magnetic
field close to the BH horizon, which is determined by the rate at which
matter is falling onto the BH. Therefore a connection between the disk
luminosity (proportional to $\dot{M}$) and the jet power (proportional to
$B\propto \dot{M}$) is naturally expected.

A simple example can be discussed for the case of BZ model and spherical
accretion. The power extractable in the BZ power can be written as:
\begin{equation} P_{BZ}\simeq
\frac{1}{128}B_0^2 r_g^2 a^2 c \,\,\,\,\, \end{equation} 
where $r_g$ is of the order of the gravitational radius of the BH.
Assuming maximal rotation for the black hole ($a=1$), the critical
problem is the estimate of the intensity reached by the magnetic field
threading the event horizon, which must be provided by the accreting
matter.  Using a spherical free fall approximation with $B_0^2/ 8\pi
\simeq \rho c^2$ one can write:
\begin{equation} P_{BZ}\simeq g \dot{M}c^2 \,\,\,\,\, \end{equation}
where $P_{acc}=\dot{M}c^2$ is the accretion power and $g$ is of order 1/64
in the spherical case, but in fact it is a highly uncertain number since
it also depends on the field configuration. 

Estimates of the power of jets and of the associated accretion flows can
therefore be crucial to shed light on the jet-disk connection .  In a
pioneering work Rawling \& Saunders (1991) addressed this question
studying a large sample of radio galaxies,
finding a good correlation between the two.  Celotti, Padovani \&
Ghisellini (1997) first investigated the relation between jet and disk
using the direct VLBI radio emission of the jet close to the nucleus for
the jet power estimation. They found a suggestive hint of correlation
between these two quantities, although the statistical significance was
too low to draw a firm conclusion.

\begin{figure}
\centerline{\hspace{4.5 cm}\epsfig{figure=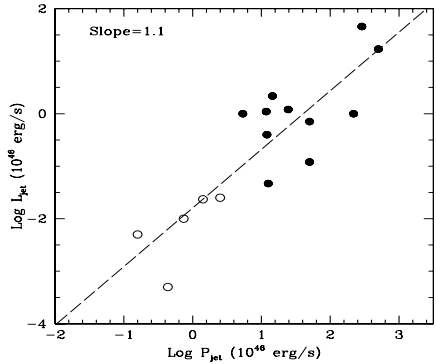, width=6.0cm, height=5.0cm}}
\caption{Radiative luminosity vs. jet power for the sample of
Blazars discussed in the text. The dashed line indicates the
least-squares fit to the data.} 
\end{figure}

\begin{figure}
\centerline{\hspace{4.5 cm}\epsfig{figure=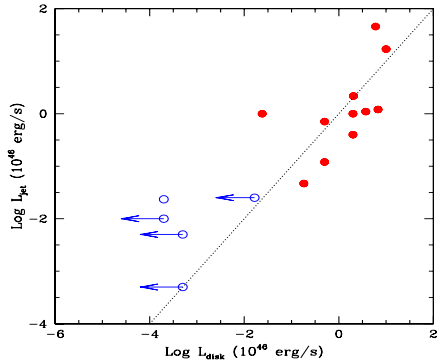, width=6.0cm, height=5.0cm}}
\caption{Radiative luminosity of jets vs disk luminosity. The line
indicates the equality between the two.}
\end{figure}

For most of the sources considered in the previous section we have also
information on the amount of radiation released by the accretion
processes, estimated either through the observed emission lines or, in
few cases, from the observed {\it blue bumps}, direct signature of the
disk. For BL Lac sources we can only infer upper limits on the disk
luminosity, but for the case of BL Lac itself, which shows a bright
line. In Fig. 2 we compare the luminosity of the jet, $L_{\rm jet}$,
which is a {\it lower limit} to $P_{\rm jet}$, with the luminosity of the
disk, $L_{\rm disk}$.  A first important result is that on average the
minimal power transported by the jet is {\it of the same order} as the
luminosity released in the accretion disk. More precisely the data we
have used suggest $L_{jet} \simeq L_{disk}$ at high luminosities and
$L_{jet} > L_{disk}$ at intermediate and low luminosities.

Assuming that $\eta = 10^{-1}$ as estimated above and that the accretion
takes place with the standard radiative efficiency $\epsilon\simeq 0.1$,
the near equality of $L_{jet}$ and $L_{disk}$ then requires $P_{\rm
jet}=P_{\rm acc}$.

On the other hand, a dominance of $L_{jet}$ over $L_{disk}$ at lower
luminosities could be attributed to a lower value of $\epsilon << 0.1$
which may be expected if the accretion rate is largely sub-Eddington.  In
the latter case the range in luminosities spanned by Fig. 2 should be
mainly a range in accretion rates rather than a range in black hole
masses.  For instance the minimum jet powers of three of the BL Lacs in
Fig 2 are around $10^{44}$ erg/s which suggests $P_{jet}\simeq 10^{45}$
requiring a mass of $10^7$ for critical accretion rate. Since the disk
luminosity is less than $10^{42}$, if the accretion rate is 1\% Eddington
the implied mass is again $10^9$ $M_{\odot}$.  

\section{Extended X-ray jets: Chandra results}

Although very common in the radio band, before the launch of the {\it
Chandra} satellite in 1999 only a handful of extragalactic kpc scale jets
were known to emit X-rays. Among them the bright and prominent jets in
3C273 , M87, Cen A, studied with {\it EINSTEIN} and {\it ROSAT}.  With
the superior sensitivity and, especially, spatial resolution of {\it
Chandra} numerous jets have been detected, triggering a new intense
theoretical and observational work.  Even in the source selected for the
first light, the distant ($z=0.6$) quasar PKS 0637-752, a prominent jet
has been discovered (Chartas et al. 2000).

Soon after the discovery of the X-ray jet in PKS 0637-0752 other jets
have been detected both in radio galaxies and in quasars. A recent census
(see e.g. the WEB site maintened by D. Harris\footnote{\small
http://hea-www.harvard.edu/XJET/}) reported 25 jets detected in
X-rays. Different classes of AGNs are represented in the sample: the most
numerous group is that of radio-galaxies (both FRI and FRII), but a large
fraction is composed by powerful radio-loud QSOs. Most of the jets have
also an optical counterpart, usually detected by {\it HST}.

The first problem posed by observations is the identification of the
emission mechanism responsible for the production of X-rays. Since these
jets are known to emit in radio, the first candidate mechanism is
synchrotron emission. In some cases (in particular for jets in
radio-galaxies) this interpretation is consistent with the data, but in
other several cases (especially in quasars) it fails to explain the
observational evidence. This is the case of the first jet discovered in
PKS 0637-0752. For this specific case Tavecchio et al. (2000) and Celotti,
Ghisellini \& Chiaberge (2000) proposed that the emission responsible for
the observed radiation is the IC scattering of the CMB radiation by
relativistic electrons in the jet.\\

\begin{figure}
\centerline{\hspace{4.5 cm}\epsfig{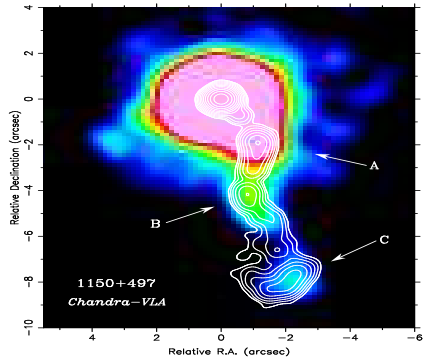}} 
\caption{{\it Chandra} (colours) and VLA (contours) images of
the jet of the quasar 1150+497 (from Sambruna et al. 2002).}
\end{figure}

In order to study in a more systematic way the properties of the X-ray
jets we proposed a series of combined {\it Chandra-HST} observations of a
selected group of quasar known to have radio jets. We also started a
program to image these jets uniformly at radio wavelengths using the VLA,
VLBA, and MERLIN. The work is still in progress. Details for a first
group of selected sources can be found in Sambruna et al. (2001). An
example is reported in Fig. 3.

\begin{figure}
\centerline{\hspace{4.5 cm}\epsfig{figure=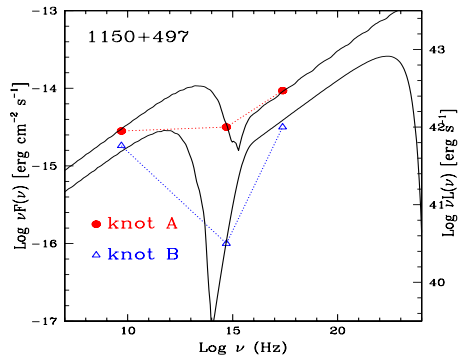, width=6.2cm, height=5.0cm}}
\caption{SEDs of the first two knots of the jet of the quasar 1150+497
with the synchrotron-IC/CMB model fit (from Sambruna et al. 2002).}
\end{figure}

\section{From the Blazar region to the kpc scale}

The possibility to constrain the physical state of the plasma in the jets
both at Blazars and kpc scale could offer us the interesting opportunity
to shed some light on the evolution of the jet from very small scales,
close to the central engine, to the outer regions, where the jet is
starting to significantly decelerate. In fact we are starting to do this
work for a small subset of the {\it Chandra} jets for which good data for
both regions are available (Tavecchio et al., in prep). 

A first important result is that the powers independently derived for
both regions are very close, supporting the overall metodology and
suggesting the idea that the jet does not substantially dissipate its
power until its end.
Another interesting suggestion derives considering the internal pressure
of the knots derived with our modelling. The derived values are in fact
close to $10^{-11}$, suggestively close to the pressure of typical hot
gas halos found around FRIs (Worrall et al. 2000). 

\begin{figure}
\vskip -0.7 true cm
\centerline{\hspace{3.5 cm}\epsfig{figure=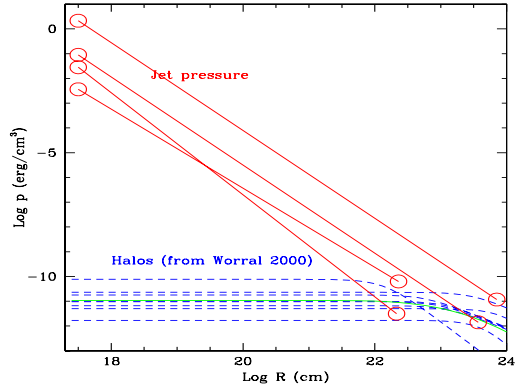, width=7cm, height=5.5cm}}
\vskip -0.3 true cm
\caption{Pressure profiles for the jet discussed in the text
(solid red lines) and for the hot halos around a group of radio galaxies
(dashed blue lines from Worrall \& Birkinshaw 2000). The green line
represents a logarithmic average of the  profiles.}
\end{figure}

The results discussed above seems to point toward a simple scenario for
the dynamical evolution of the jet from inner to the outer
regions. Clearly the pressure in the blazar jet is larger than any
possible confining gas. This strongly suggests that, after the
dissipation region, the jet expands freely until the internal pressure
(gas and/or magnetic) reaches the pressure of the (supposed) hot gas of
the galactic halo. At this point a reconfinement shock forms (the jet is
highly supersonic), particle are accelerated, and the jet starts to
decelerate (Sanders 1983).



\end{document}